Transport and Magnetic Properties of $R_{1-x}A_xCoO_3$
(R=La, Pr and Nd; A=Ba, Sr and Ca)

Hiroyasu Masuda, Toshiaki Fujita, Takeshi Miyashita, Minoru Soda, Yukio Yasui,
Yoshiaki Kobayashi and Masatoshi Sato

*Department of Physics, Division of Material Science, Nagoya University*
*Furo-cho, Chikusa-ku, Nagoya 464-8602*

**Abstract**

Transport and magnetic measurements have been carried out on perovskite Co-oxides $R_{1-x}A_xCoO_3$ (R=La, Pr, and Nd; A=Ba, Sr and Ca; $0 \leq x \leq 0.5$: All sets of the R and A species except $Nd_{1-x}Ba_xCoO_3$ have been studied.). With increasing the Sr- or Ba-concentration $x$, the system becomes metallic ferromagnet with rather large magnetic moments. For R=Pr and Nd and A=Ca, the system approaches the metal- insulator phase boundary but does not become metallic. The magnetic moments of the Ca-doped systems measured with the magnetic field $H$=0.1 T are much smaller than those of the Ba- and Sr-doped systems. The thermoelectric powers of the Ba- and Sr-doped systems decrease from large positive values of lightly doped samples to negative ones with increasing doping level, while those of Ca-doped systems remain positive. These results can be understood by considering the relationship between the average ionic radius of $R_{1-x}A_x$ and the energy difference between the low spin and intermediate spin states. We have found the resistivity-anomaly in the measurements of $Pr_{1-x}Ca_xCoO_3$ under pressure in the wide region of $x$, which indicates the existence of a phase transition different from the one reported in the very restricted region of $x \sim 0.5$ at ambient pressure [Tsubouchi et al. Phys. Rev. B 66 (2002) 052418.]. No indication of this kind of transition has been observed in other species of R.

corresponding author : M. Sato (e-mail : msato@b-lab.phys.nagoya-u.ac.jp)

# 1. Introduction

Perovskite oxides $RCoO_3$ (R=Y and various rare earth elements) have $Co^{3+}$ ions, which have the nonmagnetic ground state with the electronic configuration $t_{2g}^6$. This low spin (LS; spin $s$=0) state often exhibits a spin-state change to the intermediate spin (IS; $s$=1; $t_{2g}^5 e_g^1$) or the high spin (HS; $s$=2; $t_{2g}^4 e_g^2$) state with increasing temperature $T$.[1-11] For example, the spin-state of $LaCoO_3$ exhibits a gradual crossover with increasing $T$ from LS to IS around 100 K, and then, to HS state around 550 K.[1-6] The existence of this spin state change indicates that the energy difference $\delta E$ between the LS and IS states is rather small. Because $\delta E$ depends on the crystal field splitting $\Delta_c$ and because $\Delta_c$ generally increases with decreasing volume of the $CoO_6$ octahedra, we can control $\delta E$ by choosing the ionic radius of $R^{3+}$, though other kinds of effects may also be introduced by the radius change. The increasing tendency of the temperature of the spin state change with decreasing R-ion size[7] indicates that the idea is essentially correct.

For $R_{1-x}A_xCoO_{3-\delta}$ (A=Ba, Sr and Ca), we have to consider, at least, two kinds of effect. One is the hole-doping effect on the transport and magnetic properties. Another kind of effect is the change of $\delta E$ induced by the volume change of the $CoO_6$ octahedra or therefore by the change of $\Delta_c$. The latter induces changes of the distribution of the spin states, that is, the distribution of the hole-carriers in the $t_{2g}$ and $e_g$ orbitals changes. Effects of the double exchange interaction on the transport and the magnetic properties also change by the change of $\delta E$. Then, by controlling various kinds of parameters, species of R and A atoms, the doped hole concentration $x$, external pressure $p$ and the electron transfer energy $t$, variety of physical behaviors may be realized. The existence of the degrees of freedom of $t_{2g}$ and $e_g$ orbitals in the IS state also induces the additional variety of the physical behaviors.

Experimental studies until now, have been mainly carried out for A= Sr and Ba,[12-18] where $R_{1-x}A_xCoO_{3-\delta}$ with very small $\delta$, has been found to become ferromagnetic and metallic with increasing the doping concentration $x$.[12-15] If the R ion is relatively small and if $\delta$ is not small, oxygen-deficient perovskite structure is formed.[17,18] For example, $TbBaCo_2O_{5.5}$ has a structure in which Tb and Ba and oxygen vacancies are ordered. In this system, the metal-insulator (M-I) transition has been observed at $T=T_{M-I}\sim 350$ K.

In the present paper, we report results of measurements of the transport and magnetic properties of $R_{1-x}A_xCoO_{3-\delta}$. (R= La, Pr and Nd; A= Ba, Sr and Ca; $\delta\sim 0$. All sets of the R and A species except $Nd_{1-x}Ba_xCoO_3$ have been studied.) and discuss how the choice of A atom species affects the physical properties observed in the region $0\leq x\leq 0.5$. A M-I transition found in $Pr_{1-x}Ca_xCoO_3$ under high pressure of $p>5$ kbar is also presented.

# 2. Experiments

Polycrystalline samples of $R_{1-x}A_xCoO_{3-\delta}$ (R=La and Pr and A=Ba, Sr and Ca., $x$ = 0, 0.1, 0.2,

0.3, 0.4 and 0.5. For R= Nd and A=Sr and Ca, $x$=0, 1/8, 1/4, 3/8 and 1/2.) were prepared by the following method. For A=Ba, the samples were prepared by standard solid-state reaction. $R_2O_3$, $BaCO_3$ and $Co_3O_4$ were mixed with the proper molar ratios and the mixtures were pressed into pellets and calcined at about 1150 °C for 24 h in flowing oxygen and cooled at a rate of -100 K/h. For A=Sr and Ca, $R_2O_3$, $ACO_3$ and $Co(CH_3COO)_2 \cdot 4H_2O$ with the proper molar ratios were dissolved in the mixture of nitric acid and water. The solutions were dehydrated at 120 °C and burned at 450 °C. From the obtained powder samples, pellets with about 10 mm diameter were prepared and sintered at 900-1150 °C for 24 h under flowing oxygen and cooled at a rate of -100 K/h.

The obtained samples were annealed in high pressure oxygen atmosphere ($p$~60atm) at 600 °C for 1 day. Single-phase samples of $R_{1-x}A_xCoO_{3-\delta}$ were obtained, as confirmed by X-ray diffraction. The $\delta$ values of the obtained pellets were determined by the thermo gravimetric analysis (TGA) to be $0 \leq \delta < 0.02$. Magnetic susceptibilities $\chi$ were measured by using a Quantum Design SQUID magnetometer under the magnetic field $H$ of 0.1 T in the temperature range of 5-300 K. Electrical resistivities $\rho$ were measured by the standard four-terminal method by using an ac-resistance bridge. In the resistivity measurements, the pressure was generated with the fluid pressure medium of Daphne Oil (Idemitsu) in a clamp cell. The pressure decreases by about 2 kbar when the temperature is lowered from room temperature to 150 K. Below 150 K, the pressure is nearly $T$-independent. Further details of the measurements under pressure can be found in ref. 19. Thermoelectric powers $S$ were measured by a DC method, where the typical temperature difference between both ends of samples was about 1 K.

**3. Experimental Results and Discussion**

In Fig. 1, the cell volumes per Co determined for $R_{1-x}A_xCoO_3$ by the X-ray diffraction studies at room temperature are plotted against $x$. In the determination of the lattice parameters, we have assumed that the systems are rhombohedral for $La_{1-x}A_xCoO_3$ and orthorhombic for $Nd_{1-x}A_xCoO_3$ and $Pr_{1-x}A_xCoO_3$ though it was difficult in the present experimental resolution to distinguish if the former system is rhombohedral or cubic for $x \geq 0.3$ and if the latter two systems are orthorhombic or tetragonal for $x > 0.2$. For A= Ba and Sr, whose ionic radii are larger than those of all R species,[20] the volumes increase with $x$, while for A=$Ca^{2+}$, the volume stays almost constant in the whole $x$ region studied here, because the ionic radius of $Ca^{2+}$ is, roughly speaking, nearly equal to those of $Pr^{3+}$, $Nd^{3+}$ and $La^{3+}$.

Figure 2 shows the electrical resistivities $\rho$ of $R_{1-x}A_xCoO_3$. $R_{1-x}Sr_xCoO_3$ becomes metallic with increasing $x$ for all the present R atom species. The Ba-doped systems also exhibit the essentially metallic behavior for $x \geq 0.3$, though the slight increase of $\rho$ is observed at low temperatures with decreasing $T$ possibly due to the electron localization effect. On the other hand, the Ca-doped systems with R=Pr and Nd are nearly metallic for $x \geq 0.3$. But they seem to stay within the insulating region. The resistivity of $La_{0.5}Ca_{0.5}CoO_3$ begins to exhibit, with decreasing $T$, the metallic $T$-dependence below the temperature ~170 K, which seems to correspond to the Curie temperature of this sample(see Figs. 3 and 7), and with further decreasing $T$, $\rho$ exhibits the upturn, indicating the nature of the electron localization. The $\rho$ value of the present $La_{0.5}Ca_{0.5}CoO_3$ at room temperature is slightly larger than that reported by Muta et al.[21] It is also larger than that of the epitaxial film studied by Samoilov et al.[22]

The temperature dependence of the magnetizations of $R_{1-x}A_xCoO_3$ measured with $H$=0.1 T is shown in Fig. 3, where the data taken under the zero-field-cooling(ZFC) condition have smaller values in the low temperature region than those taken under the field-cooling(FC) condition.

As is found from the FC data, the significant spontaneous magnetic moment appears in all the Sr- and Ba-doped samples shown here, while for A=Ca, the moments are very small as compared with those for A=Ba and Sr. The hysteretic behavior of the magnetization $M$ may indicate that the systems are in the glassy phase,[14] though the possibility that the behavior is caused by the large magnetic anisotropy cannot be ruled out only by the present studies. The magnetizations of $Pr_{0.5}A_{0.5}CoO_3$ estimated by the extrapolation of the high field part of the $M$-$H$ curves to $H$=0 are ~1.2$\mu_B$/Co for A= Ba, ~1.4$\mu_B$/Co for A= Sr and ~0.12$\mu_B$/Co for A= Ca, where the $M$-$H$ taken with decreasing $H$ from 5.5 T.

In the $\rho$-$T$ curves of the metallic samples of $R_{1-x}Ba_xCoO_3$ and $R_{1-x}Sr_xCoO_3$, there exists the anomalous increase of $d\rho/dT$ with decreasing $T$ at around the temperature where the spontaneous magnetization appears, though it may not be very clear in Fig. 2, because the logarithmic scale is used there. We may also see the similar behavior of the $\rho$-$T$ curves for the samples of $La_{1-x}Ca_xCoO_3$ with relatively small resistivities($x \geq 0.3$). (The anomalous increase of $d\rho/dT$ with decreasing $T$ is clearly observed in the data shown in Fig. 7.) These anomalies can be understood by considering the Hund coupling of mobile electrons excited to the $e_g$ orbitals, with spins in the $t_{2g}$ orbitals. The coupling induces the double exchange interaction between the spins, which is the main origin of the ferromagnetic order of the systems.

In Fig. 4, the thermoelectric powers $S$ are shown against $T$. For the samples of $R_{1-x}Ba_xCoO_3$ and $R_{1-x}Sr_xCoO_3$, $S$ decreases from the large and positive values as the resistivity $\rho$ decreases with $x$, and negative $S$ is realized with increasing $x$ in their metallic phases, while for $R_{1-x}Ca_xCoO_3$, $S$ is positive in the whole $x$ region studied here. It seems to correspond to the result that the metallic state of $R_{1-x}Ca_xCoO_3$ is not observed.

In Fig.5, the magnetic susceptibilities $\chi(x)$ of $Pr_{1-x}Sr_xCoO_3$ measured under the ZFC condition are shown by the open circles in the form of $1/\chi(x)$-$T$. The susceptibilities of the Co moments, $\chi_{Co}(x)$ in $Pr_{1-x}Sr_xCoO_3$ estimated by subtracting the susceptibility $\chi_{Pr}$ of the $Pr^{3+}$ moments from the raw data are also plotted in the same form by the closed circles. In the estimation of $\chi_{Pr}$, we have first estimated the number of $Co^{3+}$ ions excited to the IS state($s=1$) in $PrCoO_3$ by using the $\delta E$ value of 1100 K and by considering the orbital degeneracy of the IS state. (The $\delta E$ value was roughly deduced from the $T$-dependence of the longitudinal nuclear relaxation rate $1/T_1$.) Then, $\chi_{Co}(0)$ was calculated for these spins by using the Curie law and $\chi_{Pr}=\chi(0)-\chi_{Co}(0)$ was obtained. For $x$ larger than 0.3, $1/\chi_{Co}(x)$ is nearly linear in $T$ above the ferromagnetic transition temperature $T_C$, where the Curie constants estimated from the lines suggest that both $Co^{3+}$ and $Co^{4+}$ are in the intermediate spin state. In the region of $x \leq 0.2$, $1/\chi_{Co}$-$T$ is nonlinear above $T_C$. This nonlinear behavior is considered to be caused by the IS→LS change of $Co^{3+}$ or both of $Co^{3+}$ and $Co^{4+}$ with decreasing $T$ for the following reasons. It is well known that $Co^{3+}$ in $PrCoO_3$ has the LS ground state and therefore the magnetic susceptibility of the system can roughly be explained by the contribution of $Pr^{3+}$ ions at relatively low temperatures. With the doping of $Sr^{2+}$ ions larger than the host ions, the crystal field energy splitting $\Delta_c$ decreases, and as the result of this decrease, the energy difference between the LS and IS states, $\delta E$ decreases. Then, the temperature of the IS→LS change becomes smaller with increasing $x$. Judging from the results shown in Fig. 5, almost all Co ions in $Pr_{1-x}Sr_xCoO_3$ are in the IS state for $x \geq x_c=0.3$ above the ferromagnetic ordering temperatures. Similar $x$ dependence of the $\chi$-$T$ curves with slightly different values of $x_c$ can be seen in other Sr-doped systems of $R_{1-x}Sr_xCoO_3$(R=La and Nd) and $R_{1-x}Ba_xCoO_3$(R=La and Pr). In contrast to these results, the nonlinear behavior is observed in the whole region of $x$ in $Pr_{1-x}Ca_xCoO_3$ and $Nd_{1-x}Ca_xCoO_3$. It is because the ionic radius of $Ca^{2+}$ is not larger than the lanthanide atoms. (For $La_{1-x}Ca_xCoO_3$, the nonlinear behavior is observed in a very restricted region of $x \leq 0.1$. Muta et al.[21] have reported that the Curie-Weiss behavior is observed in the whole region of $x$ between 0.1 and 0.5. In order to explain this slight difference, we may have to consider the possible inhomogeneity of the Ca distribution on the present samples, because the $T_C$ value of the present $La_{1-x}Ca_xCoO_3$ sample is higher than that reported in ref. 21.)

When the number of electrons in the $e_g$ orbitals or the number of Co ions in the IS state becomes significant with doping of the large ions of $Sr^{2+}$ or $Ba^{2+}$, these electrons are expected to mainly contribute to the thermoelectric power $S$, because the conductivity of the electrons is larger than that of the $t_{2g}$ holes. We expect in this case negative S. We also expect that the system is metallic and has the rather large ferromagnetic moment due to the existence of the double exchange mechanism. With the Ca doping, $\delta E$ is considered not to decrease. Then, the number of the $e_g$ electrons does not increase and only the holes introduced into the $t_{2g}$ orbitals increase. If these holes mainly contribute to $S$, its sign is expected to be positive, and because the mechanism of the double exchange interaction does not work very strongly, the magnitude of ferromagnetic moment is small in this system. These considerations can explain the rough tendency of the observed behavior of the thermoelectric powers, electrical resistivities and magnetic properties.

Now, it is interesting to realize the paramagnetic metal phase of $R_{1-x}A_xCoO_3$. At first,

measurements of the electrical resistivities ρ of $R_{1-x}Ca_xCoO_3$ in which the double exchange interaction does not become significant have been carried out under high pressure. Results for the samples of $Pr_{1-x}Ca_xCoO_3$ with $x$=0.3 and 0.5 are shown in Fig. 6. In the figure, we find the occurrence of the anomalous increase of ρ with decreasing $T$ under the pressure $p \geq 5$ kbar, which indicates the existence of a phase transition. The transition temperature increases with increasing $p$. One of interesting points is that the transition occurs for both samples with $x$=0.3 and 0.5, indicating that the transition is different from that observed at the ambient pressure by Tsubouchi et al.[23,24] for their samples of $Pr_{1-x}Ca_xCoO_3$ in the very restricted region of $x$ ~0.5 at $T$~90 K. We do not know why this difference originates. We can just say following facts: The lattice parameters of the present sample with $x$=0.5 (a=5.352(1), b=7.571(2) and c=5.367(1)) are meaningfully different from those (a=5.34045 (7), b=7.54061(9) and c=5.36766(7)) reported by the authors of ref. 23. Moreover, the present samples do not exhibit anomalous $T$-dependence of the magnetic susceptibilities at ambient pressure at around 90 K in contrast to the sample of ref. 23 with $x$=0.5. We have not succeeded in preparing samples with $x$=0.5 which exhibit, similarly to the case of ref. 23, the transition at ambient pressure.

In order to investigate if the transition found here exists in $R_{1-x}Ca_xCoO_3$ with different species of R, we have also measured the resistivities of the systems with R=La, Nd and $La_{0.2}Nd_{0.8}$. Figure 7 shows the results for $La_{0.5}Ca_{0.5}CoO_3$ and $Nd_{0.5}Ca_{0.5}CoO_3$. Although these systems exhibit gradual increase of the electrical resistivities with pressure, they do not exhibit any indication of that kind of phase transition. Figure 8 shows the results for samples of $(La_{0.2}Nd_{0.8})_{1-x}Ca_xCoO_3$ with $x$=0.3 and 0.5, where the ratio of the La and Nd numbers was chosen to adjust their average ionic radius to that of Pr. We have not observed any indication of the anomaly for these samples, either. These results suggest that the existence of the transition is characteristic of $Pr_{1-x}Ca_xCoO_3$. The present pressure induced transition does not seem to be simple charge ordering because it exists in the wide region of $x$. To clarify details of the transition, further studies including various kinds of high pressure measurements have to be carried out.

## 4. Conclusion

The transport and magnetic properties of $R_{1-x}A_xCoO_3$ (R= La, Pr and Nd; A= Ba, Sr and Ca) have been studied. With the doping of Ba and Sr into $RCoO_3$, the system becomes metallic with rather large ferromagnetic moments. For A=Ca, the system becomes nearly metallic. The ferromagnetic moments observed in the Ca-doped systems are much smaller than those of the Ba- and Sr-doped systems. The thermoelectric powers of the Ba- and Sr-doped systems become negative with increasing the doping level, while those of Ca-doped systems remain positive with the doping. These results can be understood by considering the relationship between the average ionic radius of the $R_{1-x}A_x$ and the energy difference $\delta E$ between the LS and IS states. We have found the resistivity-anomaly in the measurements of $Pr_{1-x}Ca_xCoO_3$ under pressure, which indicates the existence of the phase transition different from the one reported in the very restricted region of $x$~0.5 at ambient pressure in ref. 23. No indication of this kind of transition has been observed for other sets of R and A studied here.

Figure captions

Fig. 1  A atom concentration dependence of the volume of $R_{1-x}A_xCoO_3$ per Co atom.

Fig.2  Temperature dependence of the electrical resistivities is shown for the samples of $R_{1-x}A_xCoO_3$ with R=La, Pr and Nd.

Fig.3  Magnetizations of $R_{1-x}A_xCoO_3$ measured with the magnetic field $H$=0.1 T are plotted against $T$ for the samples with R= La, Pr and Nd, where the data taken under the ZFC condition have smaller values in the low temperature region than those taken under the FC condition.

Fig.4  Thermoelectric powers of $R_{1-x}A_xCoO_3$ are plotted against $T$ for various values of $x$.

Fig.5  Inverses of magnetic susceptibilities $1/\chi$ of $Pr_{1-x}Sr_xCoO_3$ measured with the zero-field cooling condition with $H$=0.1 T. The filled circles indicate the raw data and the closed circles show the contribution $\chi_{Co}$ of the Co moments, estimated by subtracting the contribution of $Pr^{3+}$ from the raw data.

Fig.6  Eelectrical resistivities of the samples of $Pr_{1-x}Ca_xCoO_3$ with $x$=0.3 and 0.5 taken under various values of the applied pressure.

Fig.7  Electrical resistivities of $La_{0.5}Ca_{0.5}CoO_3$ and $Nd_{0.5}Ca_{0.5}CoO_3$ taken under various values of the applied pressure.

Fig.8  Electrical resistivities of the samples of $(La_{0.2}Nd_{0.8})_{1-x}Ca_xCoO_3$ with $x$=0.3 and 0.5. The average radius of $(La_{0.2}Nd_{0.8})^{3+}$ is nearly equal to that of $Pr^{3+}$.

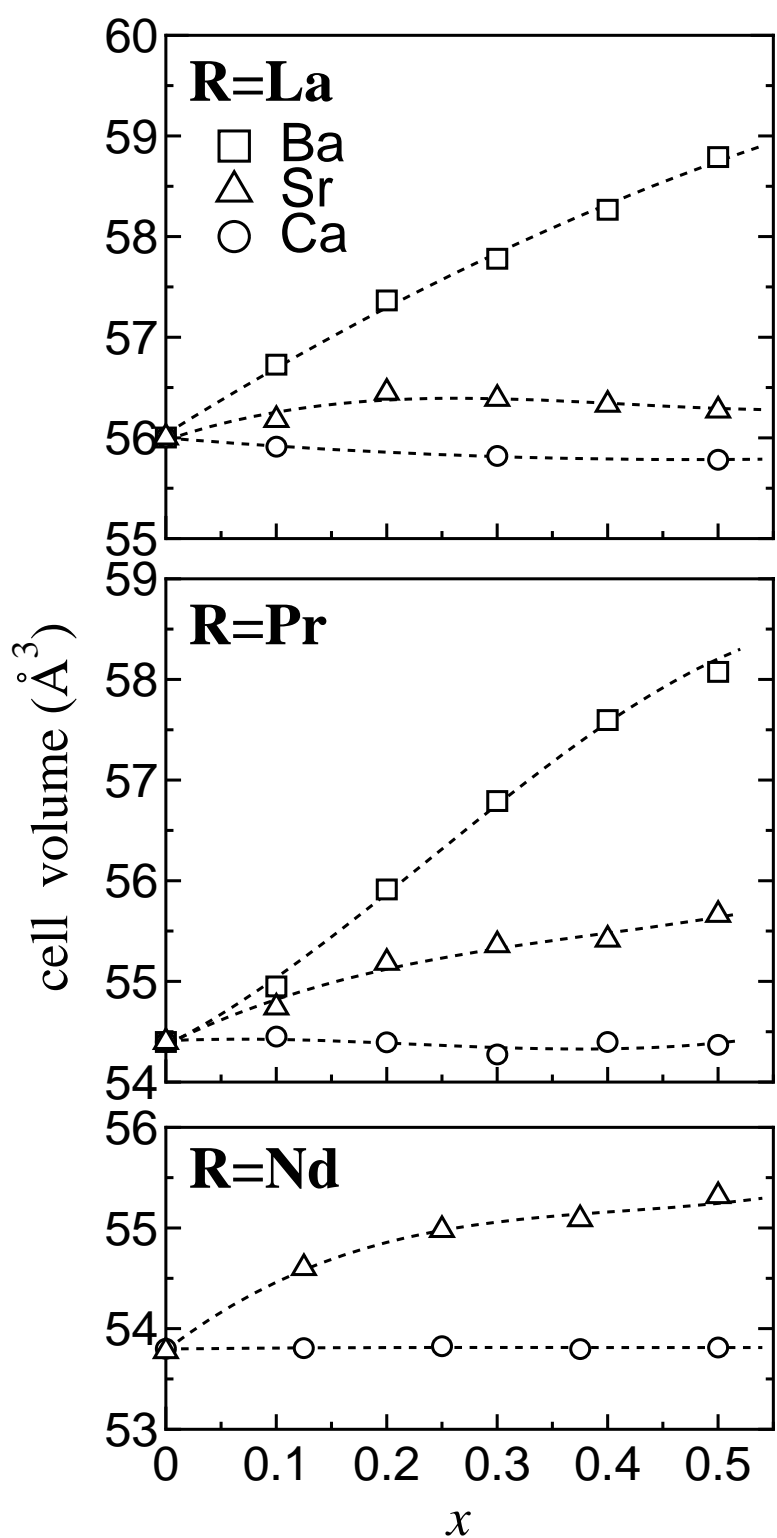

H. Masuda *et al.*

Fig.1

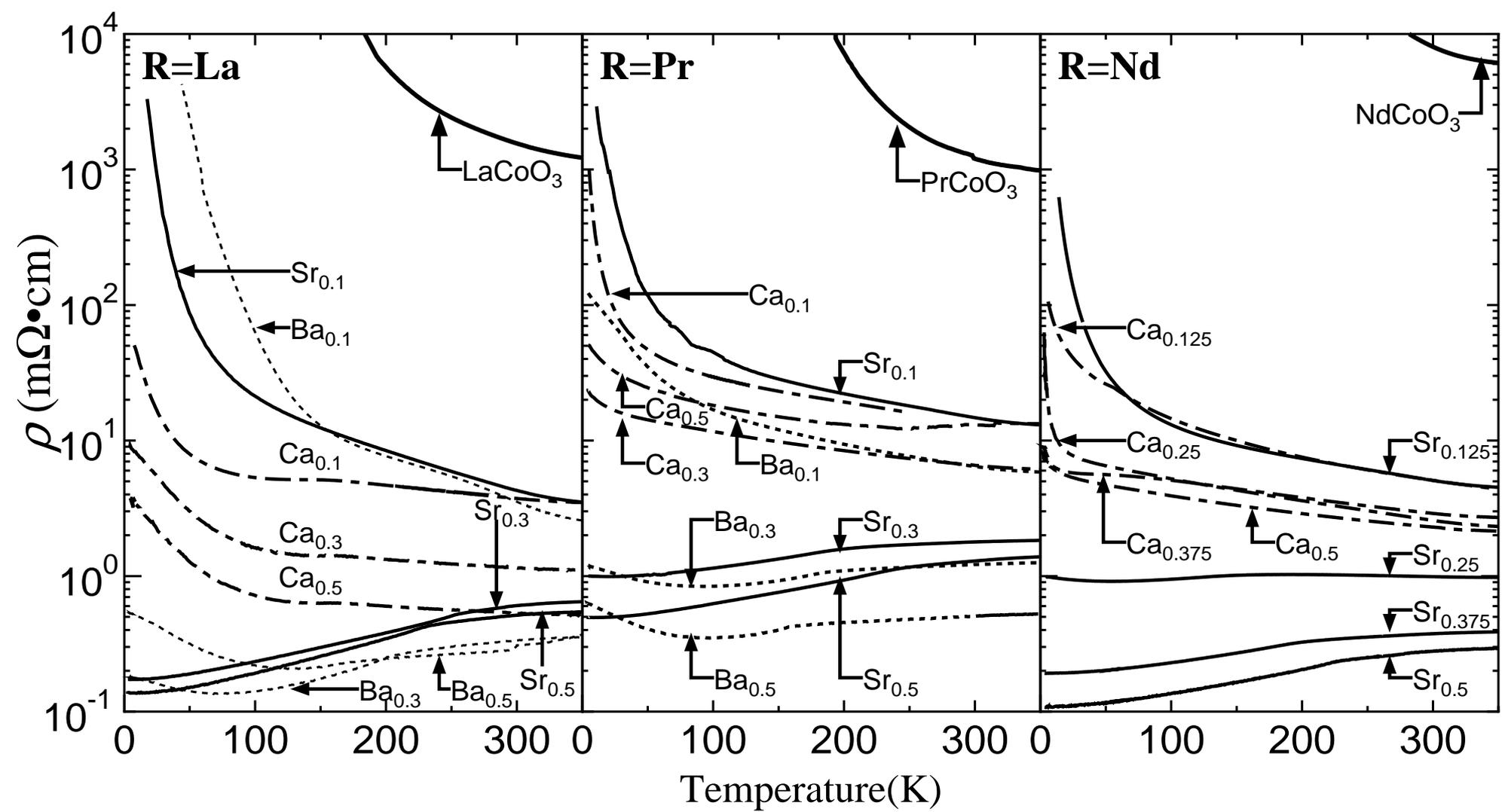

H. Masuda *et al.*

Fig.2

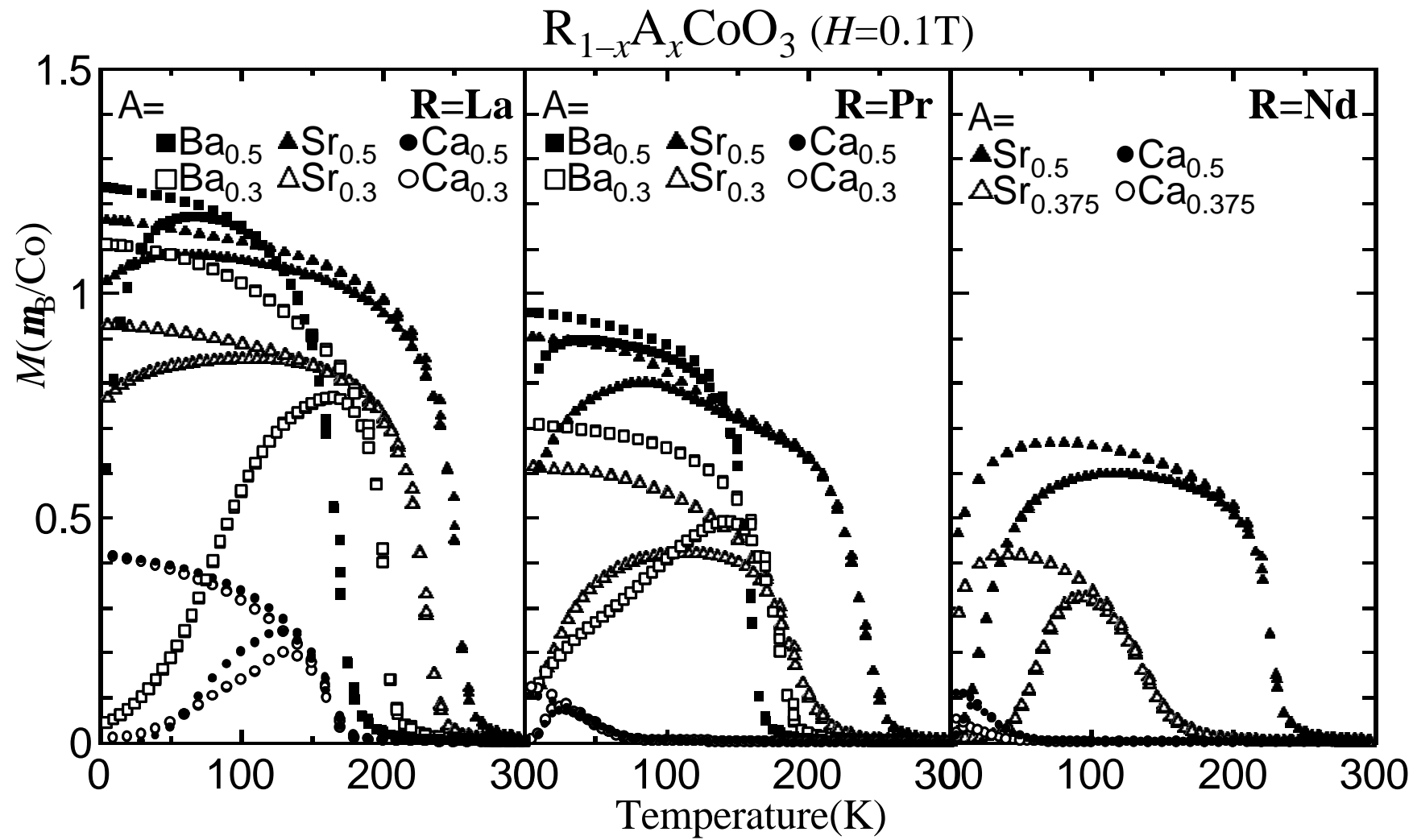

H. Masuda *et al.*

Fig.3

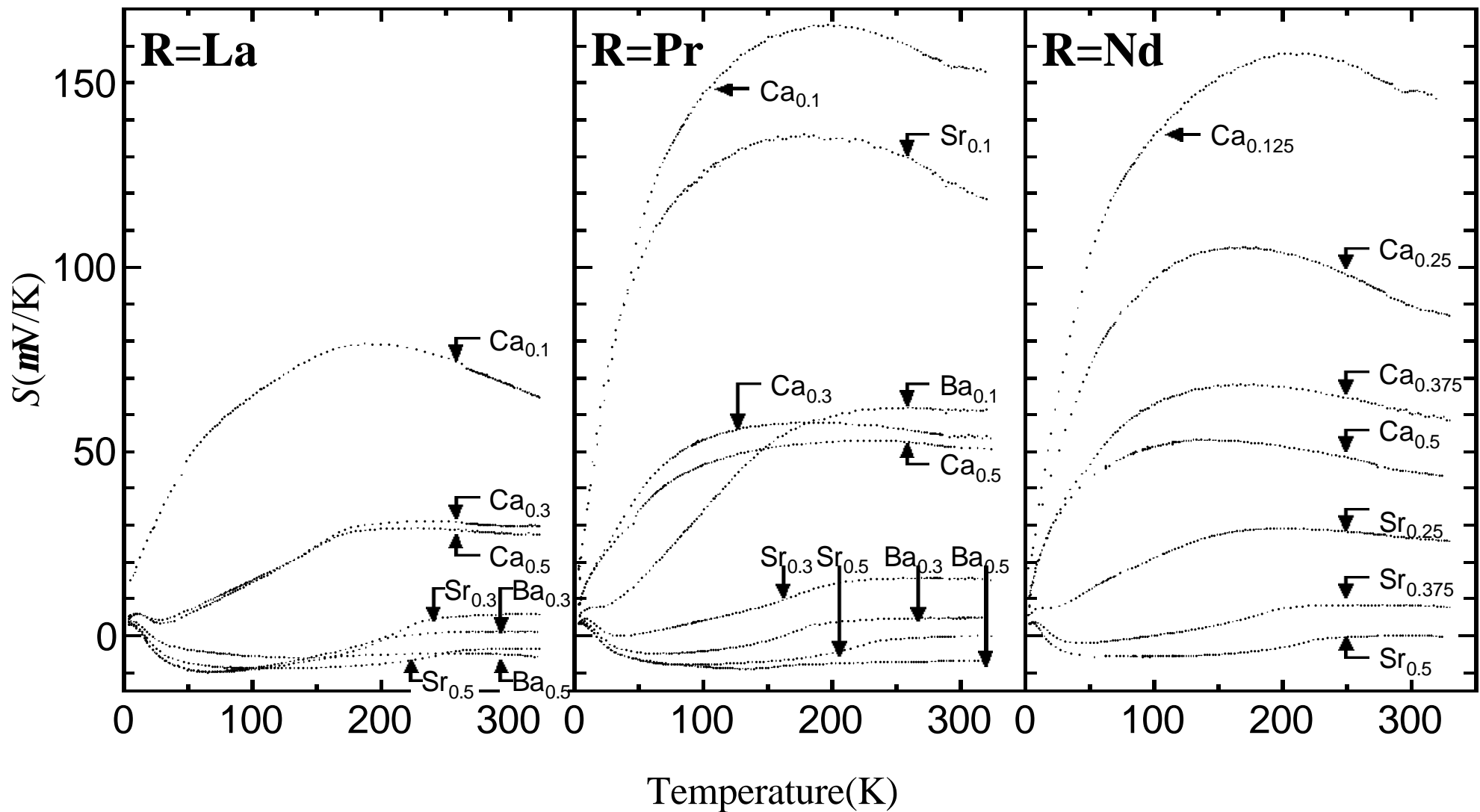

H. Masuda *et al.*

Fig.4

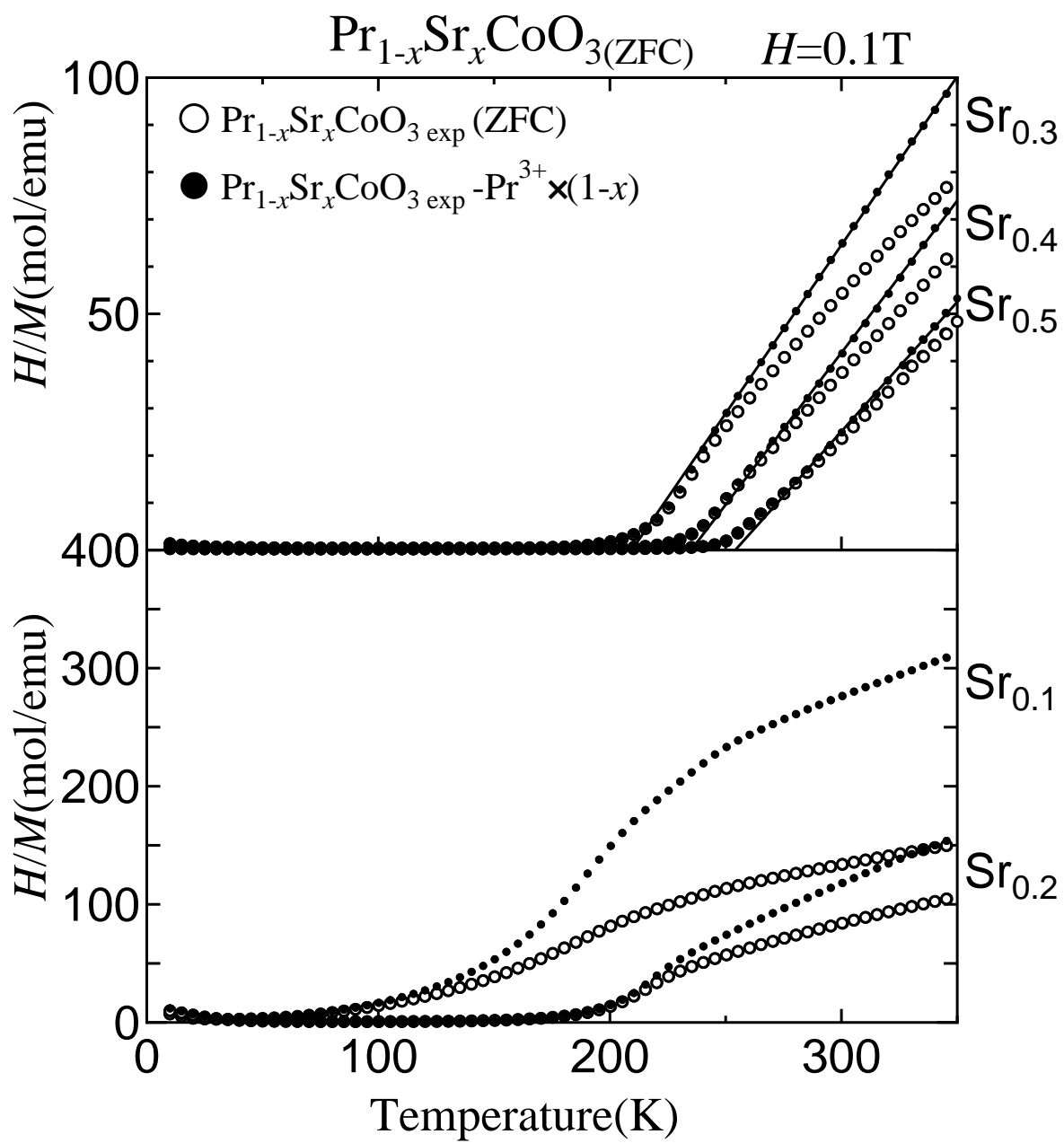

H. Masuda *et al.*

Fig.5

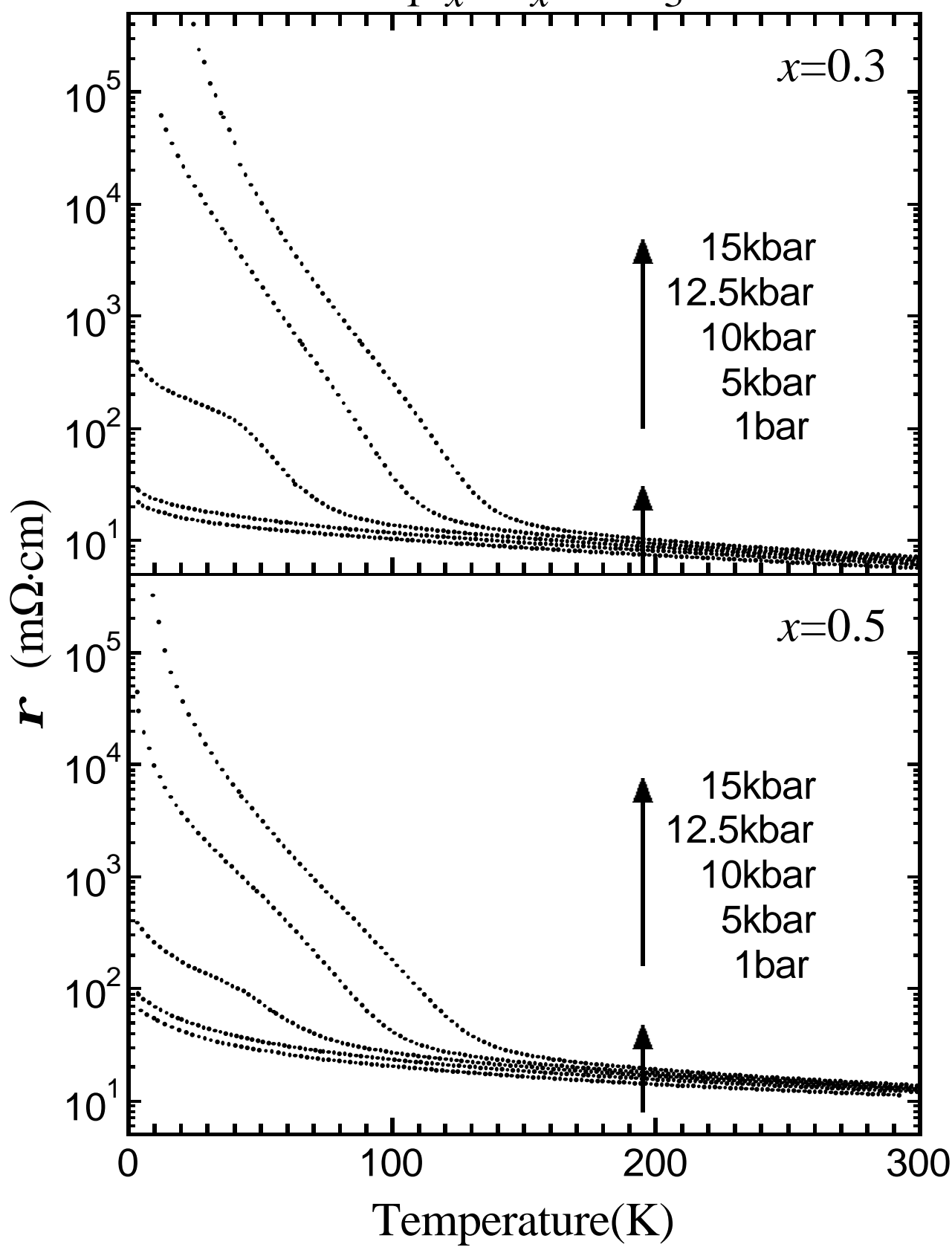

H. Masuda *et al.*

Fig.6

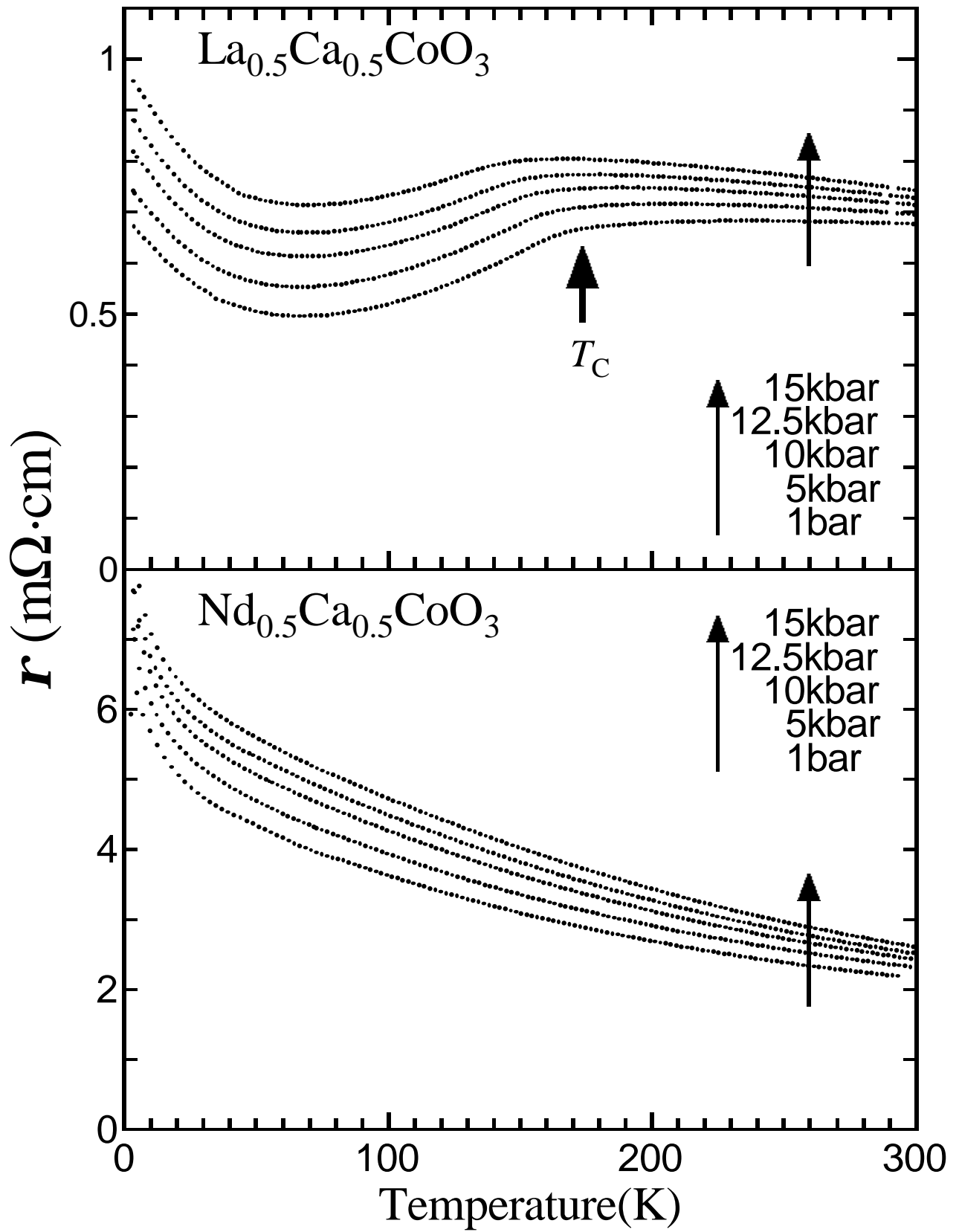

H. Masuda *et al.*

Fig.7

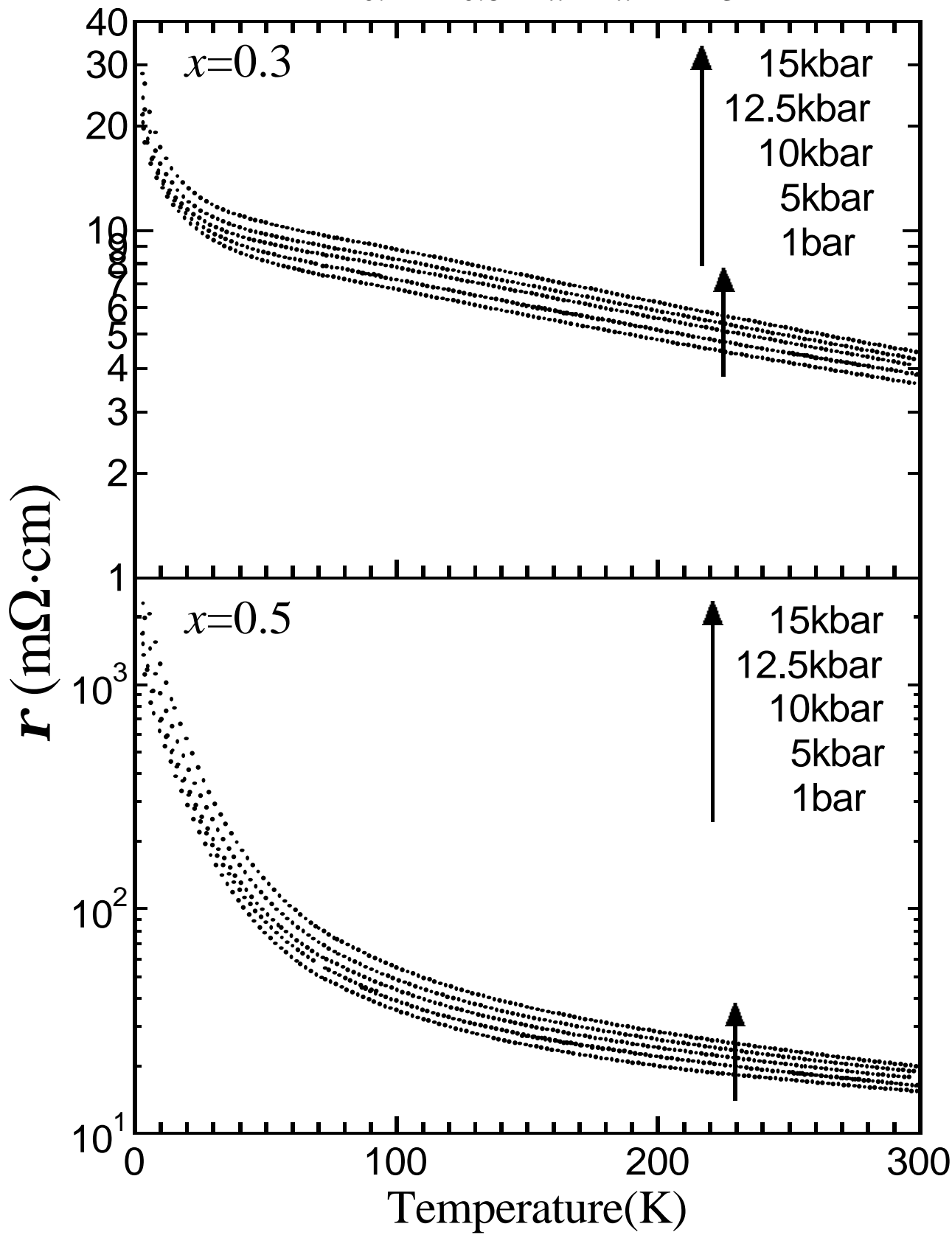

H. Masuda *et al.*

Fig.8